\documentclass[11pt,a4paper,notitlepage]{article}
\usepackage[a4paper,left=2.5cm,right=2.5cm,top=2cm,bottom=1.5cm,footnotesep=.75cm,headheight=13.6pt]{geometry}
\usepackage[center]{titlesec}
\usepackage[T1]{fontenc}
\usepackage{array}
\usepackage[utf8]{inputenc}
\usepackage{authblk}
\usepackage{nicefrac}
\usepackage{booktabs}
\usepackage[utf8]{inputenc}
\usepackage{geometry}
\usepackage{lscape}
\usepackage{tabularx}
\usepackage{graphicx}
\usepackage{rotating}
\usepackage{scalefnt}
\usepackage{multirow}
\usepackage{subfig}
\usepackage{fancyhdr}
\usepackage{float}
\usepackage{amsmath}
\usepackage{longtable}

\begin{document}

\title{Economic activity and climate change.\footnote{Financial Support from La Caixa Fundation, grant LCF/PR/SR20-52550012-Climate Change and Economic Challenges for the Spanish Society (ECHASS), is gratefully acknowledged by all the authors. Esther Ruiz also acknowledges financial support from Project PID2019-108079GB-C21 while Pilar Poncela and Arantxa de Juan are supported by Project PID2019-108079GB-C22, both from the Spanish Government. We are also very grateful to Carlos Cuerpo for invaluable comments that help us to focus this paper on relevant problems for policy makers. Any remaining errors are obviously only our responsibility.}}

\author[1]{Ar\'anzazu de Juan}

\author[1]{Pilar Poncela}
\affil[1]{Department of Economic Analysis Quantitative Economics, Universidad Aut\'{o}noma de Madrid}

\author[2]{Vladimir Rodr\'iguez-Caballero}
\affil[2]{ITAM (Mexico) and CREATES (Aarhus University)}

\author[3]{Esther Ruiz\footnote{Corresponding author. e-mail: ortega@est-econ.uc3m.es. Address: Dpto. de Estad\'istica, Universidad Carlos III de Madrid, C/ Madrid, 28903 Getafe (Spain).}}
\affil[3]{Department of Statistics, Universidad Carlos III de Madrid}

\date{14th May 2022}

\maketitle

\begin{abstract}

In this paper, we survey recent econometric contributions to measure the relationship between economic activity and climate change. Due to the critical relevance of these effects for the well-being of future generations, there is an explosion of publications devoted to measuring this relationship and its main channels. The relation between economic activity and climate change is complex with the possibility of causality running in both directions. Starting from economic activity, the channels that relate economic activity and climate change are energy consumption and the consequent pollution. Hence, we first describe the main econometric contributions about the interactions between economic activity and energy consumption, moving then to describing the contributions on the interactions between economic activity and pollution. Finally, we look at the main results on the relationship between climate change and economic activity. An important consequence of climate change is the increasing occurrence of extreme weather phenomena. Therefore, we also survey contributions on the economic effects of catastrophic climate phenomena.


\end{abstract}

\clearpage

\section{Introduction}

Measuring the interactions between economic growth and climate change could give an answer to whether it is socially beneficial for present and near future generations to sacrifice their own consumption to mitigate the latter in favour of generations yet to come and, consequently, it is critical for the design of economic and environmental policies
. As Auffhammer (2018) states, ``optimal policy design in the context of addressing the biggest environmental market failure in human history requires an understanding of the external cost imposed by additional emissions of greenhouse gases.'' In this direction, the European Union (EU) has designed a new strategy for growth directed towards its transformation into a fair and prosperous society with a modern, resource-efficient and competitive economy. As a result of this strategy, in December 2019, the EU launched the European Green Deal plan. Furthermore, as a response to the recent COVID-19 pandemic, the EU has put forward an ambitious and innovative budget for European recovery, the “Next Generation EU” with a budget of 750.000 million Euro. This plan is a game changer with respect to the discussion about climate change as member states receiving financial support should direct one third of their budgets towards the Green Economy and the ecological transition. Relevant policy recommendations need to be based on correct forecasts of the externalities caused by climate change and its mitigation costs. Robust measures of the interactions between environmental variables and the economy are crucial to formulate suitable economic development paths; see Tol (2009, 2014), who points out three of the big unknowns related to climate change, namely, extreme climate scenarios, the very long-term scenarios, and the potential economic and social effects of climate change. In this direction, the European Commission has recognized the need of designing robust quantitative systems for monitoring and forecasting key variables related with the Climate and Energy Framework objectives in order to understand whether the steps taken within each of EU members will produce the desired effects in terms of the European Green Deal plan. It is important to remark that all member states will simultaneously carry out their policies within the “Next Generation EU” plan and, by that, they will generate potential externalities that need to be measured.

Given the implications mentioned above, it is not surprising that, during the last decades, there has been an increasing interest among policy-makers, academics and the society in general, in knowing about the negative externalities of macroeconomic growth, with climate change being among the most relevant and dangerous of these externalities; see, for example, the discussions by Pindyck (2013), Stern (2008) and Stock (2019). Knowing the extent to which climate change affects important macroeconomic magnitudes, and, in particular, growth, has straightforward implications for the design of resilience economic policies. Economic growth based on burning of fossil or carbon-based fuels leads to pollution in the form of emissions of Greenhouse Gases (GHG). Higher atmospheric concentrations of GHG warm the land and oceans with impacts on temperatures, precipitation patterns, storm location and frequency, river run-off and water availability, among others. These long-run permanent changes in (short-run) weather is what is known as climate change.

The relationship between economic activity and climate change is not unidirectional as the latter may also have an impact on economic magnitudes. In a series of works, Rezai, Foley and Taylor (2012), Taylor, Rezai and Foley (2016) and Rezai, Taylor and Foley (2018) show that climate change generates externalities that may affect important economic variables as, for example, employment, income distribution and growth; see also Weitzman (2009) for a theoretical model of the economic effects of catastrophic climate change and Tanoue \textit{et al}. (2020) for the effects of floods. There are also several ``empirical'' popular models proposed to measure climate change and its economic impact. Among them, the most remarkable contribution of Nobel laureate Nordhaus is the proposal of the Dynamic Integrated model of Climate Change and the Economy (DICE), an integrated assessment model (IAM), which is a constrained non-linear dynamic optimization model with an infinite horizon; see, Nordhaus (2019) for a summary of his contributions on the economics of climate change, for which he received in 2018 the Nobel prize in Economics, and Dietz \textit{et al}. (2021) for a recent analysis of the climate policy implications of different representations of the climate system based on the DICE model. In November 2006, the British Government presented a comprehensive study based on IAMs, \textit{The Stern Review on the Economics of Climate Change}, according to which ``...if we don't act, the overall costs and risks of climate change will be equivalent to loosing at least 5\% of global GDP each year, now and forever. If a wider range of risks and impacts is taken into account, the estimates of damage could rise to 20\% of GDP or more...''; see Stern (2007) for the full text of the \textit{Review} and Nordhaus (2006) and Tol (2006) for enlighting comments on it. Barnett, Brock and Hansen (2021), Hansen (2022), Pindyck (2013, 2017), Stern (2008, 2013, 2016) and Weitzman (2007, 2015) discuss the limitations of models as DICE, which can give an illusory and misleading perception of knowledge and precision. Uncertainty is the ``Achilles' heel'' of IAMs of climate change, being of little value for a prudent current policy; see, for example, Hansen (2022), who represents the histogram of temperature projections obtained across various sets of 144 models, reflecting the substantial amount of cross-modelling uncertainty that could emerge from climate IAMs. Recently, Ikefuji \textit{et al}. (2020) propose to address some of these concerns by proposing a stochastic DICE model and introducing uncertainty about future climate change and economic changes into it.

Instead of focusing on theoretical models or on deterministic models, as those described above, in this paper, we survey over 250 empirical contributions to the literature on measuring the relationship between climate change and economic activity based on using data analysis and econometric models. The complexity of the problem makes it difficult to know where to start. Due to the huge related literature, this is a very ambitious and endless objective. Consequently, we restrict ourselves to survey the main conclusions and econometric issues faced in measuring this relationship, focusing on contributions appearing in the literature over the last decade. However, to put recent work in a broader context, in some cases, we need to briefly cover some earlier development, overlapping with some portions of available surveys. We want to apologise in advance to the many authors who have contributed to this huge literature and have not been cited in this survey.

\begin{figure}[h!]
	\centering
\includegraphics[width=0.7\textwidth]{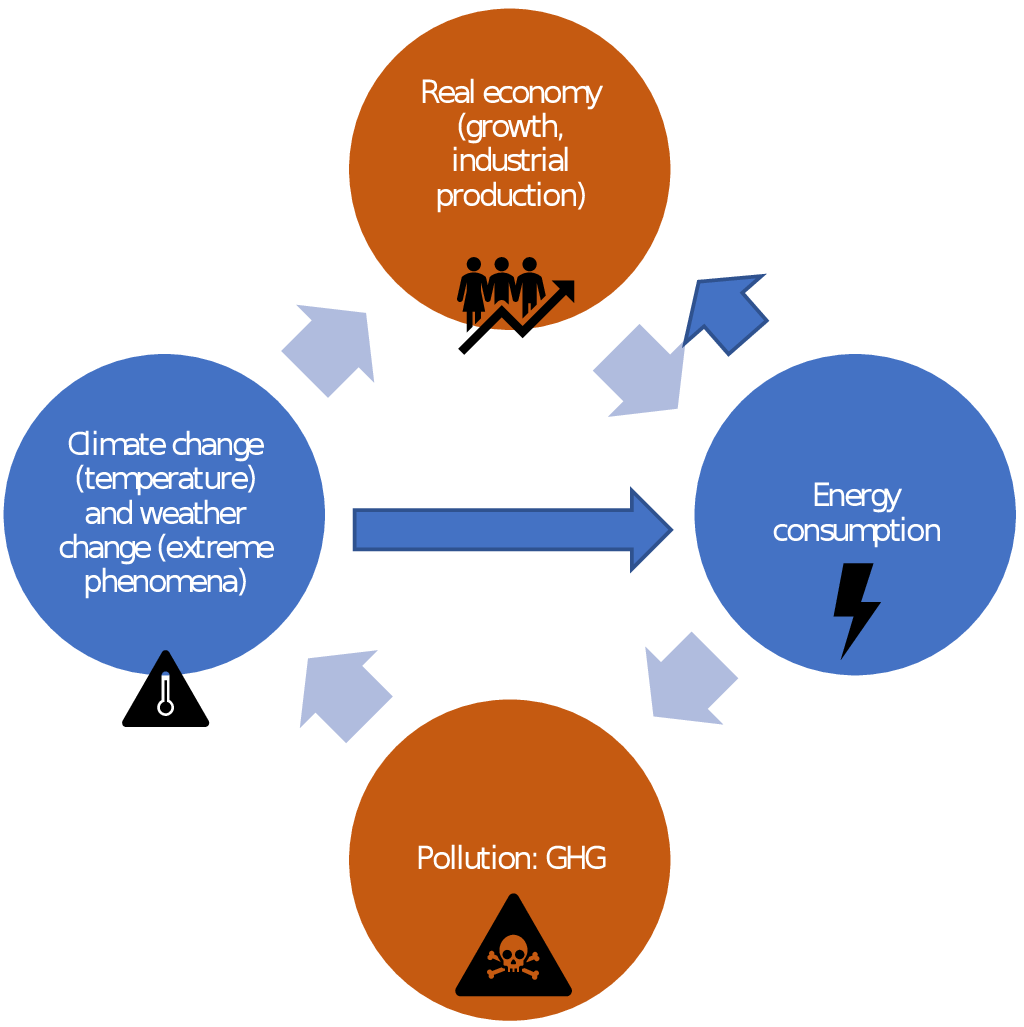}
\caption{Chart on the channels of the relationship between climate change and the economy.}
\label{fig:chart}
\end{figure}

The global picture of the channels through which climate change and the economy may be related is well understood. First, economic activity is related with an increase in energy consumption, which is related with pollution. Larger pollution is related with climate change; see, among others, 
Kaufmann, Kauppi and Stock (2006), Kaufmann \textit{et al}. (2011), Agliardi, Alexopoulos and Cech (2019), Estrada, Perron and Mart\'inez-L\'opez (2013), Chang \textit{et al}. (2020), Miller and Nam (2020) and Ero\u{g}lu, Miller and Yi\u{g}it (2022), for some evidence about the long-run relationship between anthropogenic pollution and temperatures. Finally, climate change is itself affecting the economy; see Figure \ref{fig:chart} for a chart summarizing these interactions. The first issue found when measuring and testing for the relationships described in Figure \ref{fig:chart}, is that of defining and measuring the variables involved in the different channels. The first variable that should be defined and measured is climate change itself. In the most recent literature, most measures of climate change are based on temperatures and precipitations. Given the relationship between temperatures and pollution, there is also an interest in measuring pollution, with CO$_2$ emissions being among the most popular measures because of its relevance and weight in total emissions. Finally, one should decide about the economic variables of interest to analyse the impact of climate change. In choosing the variables to focus the analysis on, one should not only choose the magnitudes to be measured but also need to decide about their spatial and temporal coverage and the frequency of observation. For example, one can carry out the analysis considering cross-sectional, panels or time-series data and, in any case, looking at global or local variables observed at different levels of aggregation and at different frequencies.

The second important issue involved in measuring the interactions between climate change and the economic activity is to decide the econometric methodology (models, estimators and tests) to be implemented, which obviously depends on the particular data under analysis; see Petris and Hendry (2013), who outline important hazards that can be encountered when modelling climate-rated time-series data and Auffhammer (2018), who describes several concerns in the empirical analysis of the economic cost of climate change. Here, we list the most relevant:
\renewcommand{\theenumi}{\roman{enumi}}
\begin{enumerate}
\item \textit{Endogeneity} due to the potential bi-directional causality between climate change and the economy. Auffhammer \textit{et al}. (2013) point out that often studies try to estimate the economic impact of climate change under a climate influenced by human activity. For example, agriculture is highly exposed to climate change because its activities directly depend on climatic conditions and, simultaneously, agriculture contributes to climate change through GHG emissions; see, Agovino \textit{et al}. (2019). 
Very recently, in a very important work, Petris (2021) discuss several issues related with the exogeneity assumption and propose a novel methodology to overcome them.
\item \textit{Non-linear relations}. For example, non-linear relationships appear in the Environmental Kuznets Curve (EKC), which postulates that the relationship between economic growth and environmental degradation follows an inverted U curve. The seminal and highly influential contribution of Grossman and Krueger (1995) concludes that there is an increase of environmental degradation and pollution in early stages of economic growth, but, beyond some level of income per capita, this relationship reverses with additional income growth leading to environmental improvement.\footnote{A number of elements related to growth as, for example, changes in the economic structure, technological progress, changes in preferences and increased environmental awareness, would be at the basis of such a relationship; see the survey by Dinda (2004), who summarizes the factors responsible for the shape of the EKC.} When the economy is mainly devoted to the agricultural activity, the environmental quality is higher than when it is mainly driven by industrial production, while when the service sector starts its development, then the environmental quality improves. The long-run relationship between economic growth and environmental degradation could thus be described as an inverted U-shaped curve, known after Kuznets (1955)\footnote{Note that Kuznets (1955) postulates the inverted U-shape for the relationship between income and income inequality.} as EKC.
\item \textit{Non-stationarity}, with potential different relations in the short or the long run; see, for example, Petris and Allen (2013), who point out that a difficulty with the statistical analysis of the relationship between temperature, GHG concentrations and the economic drives responsible for GHG changes, lies with the fact that each of these time series is non-stationary. In many works, non-stationarity is dealt with by transforming non-stationary data to stationarity. However, in doing this transformation, the information about the long run is lost; see the discussion on non-stationarity of climatic variables by Castle and Hendry (2020). Given that some of the GHGs have an atmospheric lifetime measured in tens of thousands of years, climate change is a long-run phenomenon; see Tol (2009). Therefore, measuring the long-run relationships between climate change and the economy is one the most relevant aspects of the analysis.
\item Another econometric aspect to take into account when measuring the relations between climate change and the economy is related with the potential presence of \textit{structural breaks and/or outliers}; see the discussion by Castle and Hendry (2020). To mention a recent example, consider the economic havoc observed during the present recession due to COVID-19, which will affect the state of the business cycle. The COVID-19 pandemic has caused most of the world economies to sink at unprecedented growth rates. COVID-19 has also affected the demand of energy with reductions in energy demand producing large short-run reductions in GHG emissions and uncertain long-run effects; see, for example, Gillingham \textit{et al}. (2020). 
\item An appropriate econometric model should also take into account the \textit{externalities due to the relationships among different countries}; see, for example, the results in Munir, Lean and Smyth (2020) on the between-country interactions when looking at the relationships between CO$_2$ emissions, energy consumption and economic growth. 
\item A good econometric model should take into account the \textit{uncertainty} associated with the measures of climate change and its economic implications: see Weiztman (2020), who points out that deep structural uncertainty lies at the heart of climate change economics and Calel \textit{et al}. (2020), who show that uncertainty plays a major role in computing the economic cost of climate change. Stern (2008, 2016), Pindyck (2014) and Convery and Wagner (2015) discuss that, by necessity, standard climate-economy models focus on what is known and can be quantified and, consequently, they convey a false sense of precision. Tol (2009, 2014) also argues that the level of uncertainty about the economic effects of climate change is large and understated, especially in terms of capturing downside risk. Petris \textit{et al}. (2018) remark the importance of taking into account uncertainty when looking at the economic effects of rising temperature. Finally, an appropriate measure of uncertainty is also important because, when looking at the economic impacts of climate change, analysts and policy-makers are mainly interested not in the average effect but in the left tail of the impact; see Weitzman (2009). In this context, Burke, Hsiang and Miguel (2015) and Petris and Roser (2017) point out that failing to account for climate uncertainty greatly understates the severity of the worst-case scenario.
\end{enumerate}

The rest of the paper is organized as follows. Section \ref{section:Climate} revises the measures of climate change usually used in econometric studies. Sections \ref{section:Energy}, \ref{section:Pollution} and \ref{section:Economy} are devoted to describing the main econometric contributions related to measuring the relationships between economic activity and energy consumption, energy consumption and pollution, and pollution and climate change, respectively. Section \ref{section:Disasters} reviews the economic effects of catastrophic phenomena related with weather change. Finally, section \ref{section:conclusions} concludes.   

\section{Weather variables to measure climate change}
\label{section:Climate}

In order to measure the relationships between climate change and the economy, it is crucial to start by defining and measuring climate change. This section is devoted to describing the variables often used in the empirical analysis of climate change.

Climate is usually defined as the long-run average of weather in a given location. As stated by Auffhammer \textit{et al}. (2013), "\textit{the difference between weather and climate is basically a matter of time}". The choice of weather versus climate as the variable of interest affects the interpretation of the estimated econometric models as measuring short or long-run phenomena. Hsiang and Koop (2018) define climate change as the long-run variation in the joint probability distributions describing the state of the atmosphere, oceans, and fresh water including ice.

In many studies analysing the relationships between climate and economic activity, the measures used for climate change are observations of weather variables, often temperatures and precipitations, obtained from weather stations; see Auffhammer \textit{et al}. (2013) for a discussion on some common pitfalls that empirical researchers should be aware of when using historical weather data as explanatory variables for the economy. However, one of the most important and popular aspects of climate change is Global Warming (GW), which can be described by the evolving distribution of temperature
; see the reports by Intergovernmental Panel on Climate change, IPCC (2014, 2022). Dell, Jones and Olken (2014) also point out that the second most analysed variable for climate change is precipitation. Several important works also analyse ice levels: see Castle and Hendry (2020), Diebold and Rudebush (in press) and Diebold \textit{et al}. (2021), among others. Other less popular variables are relative humidity, solar radiation, wind speed and direction and atmospheric pressure.

Given that our goal in this paper is to survey the literature analysing the relationships between climate and the economy, we focus on studies of climate data observed at the frequencies and for the time spans for which economic data are usually available, namely, monthly, quarterly or yearly data observed since the XX century.\footnote{Alternatively, many important studies of climate change consider paleo-climate observations. For example, Schmidt \textit{et al}. (2014) model paleo-climate changes for the Last Glacial Maximum, the mid-Holocene and the Last Millennium, while Castle and Hendry (2020) model climate variability over the Ice Ages. 
}

As mentioned above, one of the main characteristics of climate change is GW and, consequently, we will mainly focus on describing how GW has been modelled in the related literature. One popular methodology to obtain future projections of climate is to simulate data by General Circulation Models, also known as Global Climate Models (GCMs), which are detailed computer models that numerically approximate fundamental physical laws for representing global climate (temperatures and precipitations) in a given grid in the space; see, for example, IPCC(2014, 2022) for a description and Hsiang \textit{et al}. (2017) for a recent implementation. GCMs forecast future climate assuming heightened atmospheric concentrations of GHGs. However, there are large discrepancies across point projections of GCMs, without evidence of any model being superior than others for long-term forecasts; see, for example, Beenstock, Reingewertz and Paldor (2016) for a comparison among models and Fern\'andez \textit{et al}. (2019) for an analysis of projections made by a large number of models for the Iberian peninsula. Furthermore, the majority of GCM projections are point projections without associated measures of their uncertainty; see, Burke, Hsiang and Miguel (2015), who survey the literature on the impacts of climate change and show that the vast majority of estimates of the effects of climate on economic outcomes fail to account for the uncertainty in future temperatures and rainfall. Moreover, they argue that incorporating uncertainty into future economic impact assessments will be critical for providing the best possible information on potential impacts. Finally, Auffhammer \textit{et al}. (2013) also point out that severe biases due to spatial average can often occur when GCMs are used to simulate future climate; see Fowler, Bleukinsop and Tebaldi (2007) for bias corrections based on regressions between the GCM outputs and observed variables.

Alternatively, instead of using GCMs, many studies on GW focus on analysing the temporal evolution of particular characteristics of temperatures, hot and cold spell duration, frost days, growing season length, ice days, heating and cooling degree days, and start of spring dates; see, for example, the references cited in Diebold and Rudebush (2022). Regarding how data are collected or generated on these analysis, Dell, Jones and Olke (2014) review four types of input weather variables usually found in empirical analysis of climate change: ground station data, recorded \textit{in situ} in weather stations, gridded data, which interpolate data among ground stations, satellite data with very broad coverage although less precise and reanalysis data, which use a climate model to combine all previous types of data to enlarge the coverage of the databases. Of course, the level of temperature, i.e. the central tendency of the temperature distribution, has attracted the most attention, with many studies finding an upward trend in average daily temperatures; see, for example, Deng and Fu (2019), who compare several methods for extracting cycles from daily temperatures.

\section{Economic activity and energy consumption}
\label{section:Energy}

Kyoto Protocol and Paris Climate Accords have urged that sustainable development has been a critical part of the political climate agenda in the last decade. Thanks to climate agreements, governments have been forced to seek successful energy policies that can help to reduce environmental damage without affecting the economic growth path.\footnote{While the Kyoto protocol adopted in 1997 set top-down legally binding emission reduction targets and sanctions only for developed nations, the Paris Agreement of 2015 requires that all developed and developing countries reduce greenhouse gas emissions. Gaast (2017) describes the three main climate negotiation phases between 2005 and 2015.}

To establish successful green policies, there has been a growing interest in issues related to economic growth through energy consumption, both from a political and academic perspective. The relationship between energy and economic activity has been analysed for a long time. The early seminal short note of Kraft and Kraft (1978) introduced the necessity of studying the causal relationship between energy consumption and economic growth in the US. Since then, many papers have analysed this conundrum, including different economic mechanisms by which energy consumption and economic growth can be interlinked over time. Literature surveys by Ozturk (2010) and Payne (2010a, 2010b) provide the reader a first look at the literature using standard econometric methods on the original nexus. Tiba and Omri (2017) analyse the literature behind energy- environment-growth nexus. Waheed, Sarwar and Wei (2019) examine the survey of earlier literature that deals with economic growth, energy consumption, and carbon emission. Finally, Mutumba \textit{et al}. (2021) provide the reader with an extensive meta-analytic investigation of energy consumption and economic growth.

The goal of studying the relationship between economic activity and energy consumption is to identify their causal direction, whose consequences in the design of energy policies can be completely opposed: see the discussions by Apergis and Payne (2011) and Chen, Xie and Liao (2018) on different theories about the relationship between energy use and economic growth. The main hypotheses are described below:
\begin{itemize}
\item The Growth Hypothesis, which is supported if there exists unidirectional causality running from energy consumption to economic growth. This case is perhaps the more sensitive for energy policies due to possible direct shocks to the economic growth. This causal relationship has generated intensive participation in the global warming debate regarding the level of carbon emissions and their relationship to economic growth.
\item The Conservation Hypothesis, which demands a unidirectional causality from economic growth to energy consumption. This hypothesis would indicate to governments that reducing energy consumption would not affect the growth path and give them the breath to reach climate agreements.
\item The Neutrality Hypothesis that indicates the absence of a causal relationship between energy and growth. As in the previous case, energy conservation policies should be encouraged.
\item The Feedback Hypothesis, which requires evidence of bidirectional causality, meaning that energy consumption and growth are jointly determined. Therefore, along with the Growth Hypothesis, energy conservation policies need to be defined with the caution of avoiding to shock negatively the economic growth.
\end{itemize}

Although the literature is vast, there is still no consensus due to the heterogeneity in climate conditions, the econometric methodology employed, or even if the analysis considers only one country or a panel of them, among many other circumstances. However, as mentioned above, the common objective of this literature is clarifying the type of hypothesis supported. 

The nexus between energy consumption and economic growth has been analysed using a large variety of procedures. First, it may exhibit strong cross-sectional dependence due to the high economic and financial integration among countries and regions. Consequently, panel data models have been often used in this context; see, for example, Damette and Seghir (2013), Jalil (2014), Liddle and Lung (2015), Kais and Sami (2016) and Acheampong \textit{et al}. (2021). Many authors also consider the temporal dependence in the panel. For example, Apergis and Payne (2011) fit a panel VECM to 88 countries categorized into four panel according to their income level. They found bidirectional causality in both the short- and long-run for the high and upper-middle income countries. However, in the lower-middle income countries, the causality is unidirectional from electricity consumption to economic growth in the short run while it is bidirectional in the long-run. Finally, they find unidirectional causality from electricity consumption to economic growth for the low income countries. More recently, Lin and Benjamin (2018) also fit a fixed effect panel VECM to represent the relationship between energy consumption, foreign direct investment and economic growth for the MINT countries (Mexico, Indonesia, Nigeria and Turkey). Their conclusions about the directionality of the effects between energy consumption and growth are not uniform across all countries; see also Apergis and Payne (2009) and Kasman and Duman (2015) for applications using panel cointegration and Antonakakis, Chatziantoniou and Fillis (2017) for panel VAR.\footnote{Other proposals based on panel data take into account non-linearities as, for example, the panel quantile regression proposed by Zhu \textit{et al}. (2016)) or the nonlinear panel model proposed by Wang and Wang (2020).}

It is important to note that many papers involved in the study of the causal link between economic growth and energy consumption (even controlling for other variables such as CO$_2$ emissions) use Granger Causality (GC) as a vehicle to study the causal relationship. For example, Chen, Xie and Liao (2018) implement the panel Granger causality analysis proposed by Emirmahmutoglu and Kose (2011) to examine the causal relationship between energy consumption and energy growth for 29 Chinesse provinces. They conclude that there is a unidirectional causal link running from real output to energy use. However, such tests cannot be considered neither a necessary nor a sufficient condition for causality since it does completely characterize the notion of ``cause''; see the discussion by Hendry (2004). In contrast, to analyse the causal link between economic growth and energy consumption and further control variables, Rodr\'iguez-Caballero and Ventosa-Santaularia (2017) propose going back to the original ideas of exogeneity emboided in Engle, Hendry and Richard (1983) and study a causal link through the statistical concepts of weak and super exogeneity.

Very recently, Rodr\'iguez-Caballero (2022) proposes a non-stationary panel data model with multi-level cross-sectional dependence to study the long-run relationship between economic growth and energy consumption in 69 countries. In his model, the commonality between countries is driven by unobservable common factors, which can gather relevant variables affecting all countries such as international shocks generating financial crisis, the sharp decrease in oil prices, or by climate conditions.

Second, in the context of single-equation time series, Shahbaz \textit{et al}. (2018) propose dealing with potential non-linearities in the nexus between energy use and economic growth by using the quantile-on-quantile (QoQ) methodology recently proposed by Sim and Zhou (2015). QoQ combines quantile regression and non-parametric estimation to regress a quantile of GDP on a quantile of energy consumption and vice-versa. After fitting QoQ to top ten energy-consumption countries, they find a weak effect of economic growth on energy consumption for lower quantiles in some countries and for upper quantiles in others. Furthermore, the effect of energy consumption on energy growth is, in general, even weaker. However, it is important to point out that, although the QoQ is an interesting way of dealing with non-linearity, Shahbaz \textit{et al}. (2018) do not take into account the non-stationarity of both energy consumption and economic growth and, consequently, their results may be spurious.   

Finally, note that climate change can alter energy generation potentials and needs, adding to the route that energy-growth conundrums could follow; see Figure \ref{fig:chart}. Naturally, climate change could excessively increase the electricity demand for cooling during heat waves. New literature is emerging intending to analyse the direct impact of climate change on energy and vice versa, through renewable energy; see, for example, Cronin, Anandarajah and Dessens (2018), Quaschning (2019), Van Ruijven, de Cian and Wing (2019) and Olabi and Abdelkareem (2022). A review of climate change impacts on renewable energies is provided by Solaun and Cerd\'a (2019).

\section{Economic activity and pollution}
\label{section:Pollution}

GHG emissions are a major cause of climate change. They have a strong relationship with energy consumption, which, as described in the previous section, is closely related with economic activity. Since the seminal works of Selden and Song (1994), Grossman and Krueger (1995) and Holtz-Eakin and Selden (1995), the econometric analysis of the relationship between economic activity and pollution has been the focus of an extremely large literature; see the relatively recent surveys by Al-Mulali, Saboori and Ozturk (2015), Moutinho, Varum and Madaleno (2017), Stern (2017), Shahbaz and Sinha (2019), Waheed, Sarwar and Wei (2019), Purcel (2020) and Ul Husnain, Hider and Khan (2021), and the references therein. The number of works is so large that trying to survey this literature seems an impossible mission. Consequently, in this section, we just summarize some of the main findings of the literature on the nexus between economic activity and pollution. Most works assume that the former variable is exogenous and represent the EKC by the following static (long-run) model
\begin{equation}
\label{eq:EKC}
e_{it}=\beta _{0i}+\beta _{1i}y_{it}+\beta _{2i}y_{it}^{2}+\beta_{3i}y_{it}^{3}+u_{it},
\end{equation}
where $e_{it}$ represents the logarithmic transformation of the environmental quality proxy in country $i$ at time $t$, and $y_{it}$ is the corresponding logarithmic transformation of the economic measure. For each $i$, $u_{it}$ is assumed to be an independent white noise with variance $\sigma^2_{u_i}$. If $\beta _{1i}>0,\beta _{2i}<0$ and $\beta _{3i}=0$, then (\ref{eq:EKC}) represents an inverted-U shape in the relationship between pollutants and economic activity. Furthermore, when $\beta_{3i}\neq 0$, the relationship between pollution and per capita GDP has an N-shaped relationship (which can be inverted depending on the sign of the parameters in (\ref{eq:EKC})); see Moosa (2017) for an interesting discussion on the shape of the relationship between pollution and GDP. For a given country $i$, if $\beta_{3i}=0$, the turning point of the per capita GDP in which the environmental quality measure, $e_{it}$, begins to improve is $y_i^{\ast }=-\frac{\beta _{1i}}{2\beta _{2i}}$.

Next, we describe the variables that have been usually chosen to represent environmental quality and economic activity as well as the methods implemented to estimate model \eqref{eq:EKC}.

\subsection{Variables to measure pollution and economic activity}

With respect to the variables used to measure pollution, empirical studies of the EKC consider at least one of eight possible broad types of pollution, namely, air pollution, water pollution, land pollution, radioactive pollution, noise pollution, light pollution, thermal pollution and ocean or marine pollution.\footnote{Dinda (2004) points out that EKC cannot be generalized for all types of pollutants, being not valid, for example, for industrial water pollution or toxic pollution.} However, one of the most enduring sources of GW, and consequently of climate change, is the burning of fossil or carbon-based fuels as coal, oil and natural gas, which leads to emissions of GHG, with carbon dioxide (CO$_{2}$) being among the most important. Consequently, the analysis of the effects of economic growth on the environment has been mainly done considering air pollution, and more specifically CO$_{2}$ emissions, which remain accumulated in the atmosphere for a long time; see, for instance, the references listed in Tables \ref{tab:panel} and \ref{tab:TS} in which we summarize empirical results on the EKC and Al-Mulali, Saboori and Ozturk (2015) and Ul Husnain, Haider and Khan (2021) for similar tables. Very few empirical investigations about the EKC have been done using sulphur dioxide (SO$_{2})$ emissions or other air pollutants; see, for example, Stern and Common (2001) and Taguchi (2013). Finally, it is important to mention that, in a recent study, Altintas and Kassouri (2020) consider two indicators of environmental degradation, namely, ecological footprint and CO$_2$ emissions, and find evidence for the sensitivity of the EKC hypothesis to the type of environmental degradation proxy used and conclude about the relevance of the ecological footprint as an appropriate environmental tool. Gill \textit{et al}. (2017) and Rao and Yan (2020) also point out that the EKC has only been proved for a subset of pollutant indicators. Recently, Haider, Bashir and Ul Husnain (2020) find strong support for the EKC using NO$_2$ as pollutant and Pandey and Mishra (2021) obtain a similar result using SO$_2$ and NO$_2$ considering 21 states of India.

With respect to the variables used to represent economic activity, most works usually consider per capita income or Gross Domestic Product (GDP), although alternative explanatory variables have also been added to explain pollution. For example, Pablo-Romero, Cruz and Barata (2017) test the EKC using transport as a proxy for economic activity. Furthermore, according to the Pollutant Heaven Hypothesis, developed countries look for the cheapest options in terms of resources and labour and often set up factories abroad. In this case, even if their economic activity increases, they may not have increases in their pollutant emissions. However, the developing countries that recieve the factories may show larger emissions without having larger GDP. For this reason, many studies of the EKC control for variables related to trade; see, among others, Jebli, Youssef and Ozturk (2016), Tiwary, Shahbaz and Hye (2013), Lau, Choong and Eng (2014), Onafowara and Owoye (2014), Dogan and Turkekul (2016), Anser \textit{et al}. (2020), Adebayo, Awosusi an Adesola (2020), Khan and Eggoh (2021) and Bidi and Jamil (2021), for recent references.

Alternatively, many works explore the link between the economy and CO$_2$ emissions through the energy channel; see, inter alia, Ang (2007), Acaravci and Ozturk (2010), Shahbaz, Mutascu and Azim (2013), Smiech and Papiez (2014), Acheampong (2018), Chen, Wang and Zhong (2019), Diaz \textit{et al}. (2019), Usman, Larember and Olanipekun (2019), Munir, Lean and Smyth (2020), Hasmi \textit{et al}. (2021), Altintas and Kassouri (2020), Saidi and Omri (2020), Khattak \textit{et al}. (2020), Erdogan (2020), Raza, Shah and Khan (2020) and Cheikh, Zaied and Chevallier (2021). Other authors have also controlled for other variables as Hailemariam, Dzhumashev and Shahbaz (2020), who consider the effect of income inequality, Balaguer and Cantavella (2016), who add fuel prices as an indicator of fuel energy consumption, Hasmi \textit{et al}. (2020), who consider the effect of geopolitical risk or Usman, Larember and Olanipekun (2019), who consider the effect of democratic regime. Finally, Hip\'{o}lito Leal and Cardoso Marques (2020) analyse the repercussion of globalization on the environment.

The potential set of variables that drive emissions of CO$_2$ is very large. Furthermore, it is also important to analyse the functional form in which these variables enter the EKC; see the very extensive analysis by Auffhammer and Steinhauser (2012), who compare over 27000 specifications arising from possible permutation of a very limited set of explanatory variables. 

Finally, there are authors considering variables that are not properly economic although they are closely related. For example, Wang \textit{et al}. (2021) analyse the long- and short-run relationships between urbanization and three carbon emission dimensions in OECD high income countries. They conclude that developed countries tend to have the same negative impact of urbanization on carbon emissions, although there are differences on the endowments of different countries; see also Adebayo and Odugbesan (2021), who also consider the role of urbanization on CO$_2$ emissions. Chen, Huang and Lui (2019) consider the effects of emissions on environmental awareness. Beyene and Kotosz (2020) and Aziz \textit{et al}. (2021) analyse the effect of globalization on CO$_2$ emissions and Anset et al (2020), Aller et al (2020), Lee, Chen and Wu (in press) and Nosheen et al (2021) emphasize the effect of tourism on pollutant emissions.

\subsection{Estimators of the EKC}

In this subsection, we survey different procedures implemented to estimate the EKC, clasifying them into panel, time series and regression estimators.

\subsubsection{Panel data}

Estimation of model (\ref{eq:EKC}) is usually carried out using panel data and the fixed effects estimator; see, for example, the meta analysis carried out by Sarkodie and Strezov (2019) on the estimation of the EKC. Very recently, instead of the fixed effects estimator, Aller, Ductor and Grechyna (2021) propose using Bayesian Model Averaging and Cluster-LASSO to estimate all possible combinations of the regressors, taking afterwards a weighted average over the candidate models. They conclude that CO$_2$ emissions in high- and medium-income economies are affected by the share of industry in GDP, political polarization and tourism, while, in low-income countries, they are positively affected by foreign direct investment, the level of democracy and corruption. It is important to point out that the relationship in model (\ref{eq:EKC}) may depend on the level of economic development of the countries included in the analysis with low developed countries in the first increasing part of the inverted-U. To account for this heterogeneity, Stern and Common (2001) and Galeotti, Lanza and Pauli (2006) consider specific intercepts and found support of the EKC when applied to a panel with a large number of countries. Moutinho, Varum and Madaleno (2017) also use panel data on different sectors to investigate the EKC in Portuguese and Spanish economic activity sectors.

Estimates of model (\ref{eq:EKC}) could be affected by simultaneity biases resulting from the fact that pollutants could also cause economic activity; see the surveys by Tiba and Omri (2017) and Waheed, Sarwar and Wei (2019) on the causal directions between economic growth, energy and pollution. To control for the potential endogeneity of growth, Lin and Liscow (2013) propose an instrumental variable (IV) estimator, which is implemented to a panel of OECD and non-OECD countries. Using several measures of water pollution as pollulants, and controlling for the indices on political rights and civil liberties from Freedom House, Lin and Liscow (2013) find empirical support of the EKC hypothesis and conclude that controlling for variables representing social issues is important in this relationship.

Due to the strong persistence often observed in the variables involved in the estimation of the EKC, panel models have been estimated using procedures designed to deal with this characteristic. Acheampong (2018) proposes an integrated framework to analyse Granger causality between economic growth, energy consumption and carbon emissions, based on a panel Vector Autoregression model estimated using the system-GMM of Blundell and Bond (1998) that takes into account strong persistence. In a very recent work, Munir, Lean and Smyth (2020) propose testing for Granger causality among CO$_2$ emissions, energy consumption and economic growth using the estimator of Westerlund (2007), taking into account not only cointegration but also heterogeneity and cross-sectional dependence. They show that panel unit root and cointegration tests that do not accommodate cross-sectional dependence give mixed and inconclusive results. After implementing the model to the five main countries of the Association of Southeaster Asian Nations (ASEAN-5) over the period 1980-2016, they conclude that in three countries, Indonesia, Malaysia and Thailand, there is unidirectional causation running from GDP to energy consumption. In these countries, energy conservation is unlikely to have much impact on economic growth. For four countries, Malaysia, Philippines, Singapore and Thailand, they find unidirectional causality running from GDP to CO$_2$. They also conclude that three of these countries, Malaysia, Philippines and Thailand, have not yet reached the income turning point suggested by the EKC. For these countries, economic growth can be expected to adversely affect the environment until the turning point is reached. The Westerlund's (2007) panel cointegration method has also been used in Zafar \textit{et al}. (2020) to verify long-run relationship between CO$_2$ emissions and GDP, controlling for energy consumption, urban population and industrialization in a panel regression model for 46 countries covering the period 1980-2017. Another application of this methodology can be found in Altintas and Kassouri (2020), who analysed the EKC relationship using a heterogeneous panel for 14 European countries over the period 1990-2014. Alternatively, several authors have implemented the cointegration tests for panel data proposed by Pedroni (2004); see, for example, Erdogan (2020), Lazar \textit{et al}. (2019), Saidi and Omri (2020), Adebayo, Awosusi and Adesola (2020) and Dogan and Inglesi-Lotz (2020).

Finally, several authors analyse the EKC hypothesis using the panel extension of the Autoregressive Distributed Lag (ARDL) procedure of Pesaran, Shin and Smith (1999, 2001)\footnote{Shin, Yu and Greenwood-Nimmo (2014) propose non-linear ARDL tests. More recently, McNown, Sam and Goh (2018) propose a bootstrap correction of the ARDL tests with better properties.}; see Hanif \textit{et al}. (2019), who support the EKC for a panel of 15 Asian developing countries, observed over the period 1990 - 2013 and Waqih \textit{et al}. (2019), who also find support for the EKC in a panel of four countries of the South Asian Association for Regional Cooperation (SAARC), observed during the period 1986-2014 based on combining the ARDL\ approach with the Full Modified OLS.

Given that the channel of the relationship between economic activity and pollution is energy consumption, Marrero (2010) studies the simultaneous relationship between emissions, growth and energy based on a dynamic panel of 24 European countries observed from 1990 to 2006. He concludes that the elasticity between aggregate energy consumption and emissions is significantly greater than zero and below unity. However, he does not find evidence in favour of the EKC hypothesis.

In the context of panel data, several authors have estimated model (\ref{eq:EKC}) using procedures from spatial econometrics; see, for example, Kang, Zhai and Yang (2016), who consider a panel of Chinese provinces over the period 1997-2012 and find an inverted-N shaped relationship and Hao \textit{et al}. (2016), who consider coal consumption as the endogenous variable and find a bell shape relationship. Li and Wang (2019) also apply spatial econometrics to a panel of 30 Chinese provinces. They use carbon intensity of human well-being as the pollutant measure and find an inverted-N shaped relationship.

\begin{sidewaystable}
	\caption{Summary of selected works testing the EKC using panel data methods.}
	\resizebox{0.6\textwidth}{!}{\begin{minipage}{\textwidth}
			\begin{tabular}{cccccccccccccc}
		{\small Authors} & {\small Dependent Variable} & {\small Independent
			Variables} & {\small Countries} & {\small Sample} & {\small Conclusion} & 
		{\small Method} \\
		\hline
		{\small Adebayo \textit{et al}. (2020)} & {\small CO$_{2}$} & {\small EG, Energy use, trade, urbanization} & {\small MINT economies} & {\small 1980 - 2018} & {\small Cointegration CO$_{2}$}{\small \ and determinants} & {\small Panel Cointegration, ARDL-MMG} \\
		{\small Altintas and Kassouri (2020)} & {\small Ecological Footprint, CO}$_{2}$ & {\small Renewable energy, Fossil fuel consumption} & {\small 14
			European Countries} & {\small 1990 - 2014} & {\small Depend} & {\small %
			Heterogenous panel} \\
		{\small Anser \textit{et al}. (2020)} & {\small CO}$_{2}$ & {\small International
			tourism, social distribution, FDI} & {\small G7 countries} & {\small 1995 -
			2015} & {\small Support EKC, PHH and REH} & {\small Panel random Effect and
			Panel} \\ 
		{\small Beyene and Kotosz (2020)} & {\small CO}$_{2}$ & {\small %
			Globalization, FDI, Pop. density, Pol.stability} & {\small 12 East African
			countries} & {\small 1990 - 2013} & {\small Support EKC in the short run} & 
		{\small Pedroni's test, Pooled Mean Group} \\ 
		{\small Cai \textit{et al}. (2020)} & {\small Different pollutants} & 
		{\small EG} & {\small China} & {\small 2003 - 2017} & {\small Different
			types of EKC} & {\small Locally weighted smoothed} \\ 
		{\small Dogan and Inglesi-Lotz (2020)} & {\small CO}$_{2}$ & {\small EG,
			Economic Structure} & {\small 7 European countries} & {\small 1980 - 2014} & 
		{\small Support EKC} & {\small Panel UIR, Panel CI tests, FMOLS} \\ 
		{\small Erdogan (2020) } & {\small Ecological Footprint, CO}$_{2}$ & {\small %
			Renewable energy, fossil fuel, consumption} & {\small Europe} & {\small 1990
			- 2014} & {\small Sensitivity of the EKC to the pollutant} & {\small %
			Cross-section and slope heterogeneity, Cl} \\ 
		{\small Haider \textit{et al}. (2020)} & {\small NO}$_{2}$ & {\small Agricultural land use, exports, EC} & {\small Developed, developing countries} & {\small 1980 - 2012} & {\small Support EKC } & {\small Pooled mean group approach}\\ 
		{\small Jiang \textit{et al}. (2020)} & {\small Air pollution} & {\small EG} & 
		{\small China and South Korea} & {\small 2003 - 2017} & {\small Support EKC}
		& {\small Panel data} \\ 
		{\small Khattak \textit{et al}. (2020)} & {\small CO}$_{2}$ & {\small Innovation,
			renewable energy consumption} & {\small BRICS countries} & {\small 1980 -
			2016} & {\small Support EKC} & {\small Johansen + Fisher, Kao's CI tests }\\ 
		{\small Leal and Marques (2020)} & {\small CO}$_{2}$ & {\small Degree of
			Globalization} & {\small 20 OECD countries} & {\small 1990 - 2016} & {\small %
			Support EKC} & {\small Driskoll - Kraay estimator} \\ 
		{\small Munir \textit{et al}. (2020)} & {\small CO}$_{2}$ & {\small Energy
			consumption, EG} & {\small ASEAN-5} & {\small 1980 --2016} & {\small %
			Inconclusive results} & {\small Panel VAR} \\ 
		{\small Raza \textit{et al}. (2020) } & {\small Residential energy} & {\small EG,
			renewable energy, development} & {\small NEXT11 and BRICS countries} & 
		{\small 1990 - 2015} & {\small Support EKC } & {\small FMOLS, Pedroni
			test,Westerlung (2007)} \\ 
		{\small Saidi and Omri (2020)} & {\small CO}$_{2}$ & {\small Renewable
			energy, EG} & {\small 15 major renewable energy consuming} & {\small 1990 -
			2014} & {\small Support EKC in the short run} & {\small Pedroni's test, FMOLS%
		} \\ 
		{\small Akadin \textit{et al}. (2021)} & {\small CO}$_{2}$ & {\small EG, Economic
			freedom} & {\small BRICS countries} & {\small 1995 - 2018} & {\small Support
			EKC in the long-run} & {\small Pooled mean Group estimation} \\ 
		{\small Aller \textit{et al}. (2021)} & {\small CO}$_{2}$ & {\small \% ind. in GDP,
			Pol. polarization, Tourism, FDI} & {\small High and medium income countries}
		& {\small 1995 - 2014} & {\small Support EKC} & {\small BMA / and
			Cluster/LASSO} \\ 
		{\small Aziz \textit{et al}. (2021)} & {\small CO}$_{2}${\small \ } & {\small Nat.
			ressources, renewable energy and globalization} & {\small MINT countries} & 
		{\small 1995 - 2018} & {\small Not Support highest quantiles} & {\small %
			Pedroni's test, FMOLS, DOLS, FE-DOLS} \\ 
		{\small Bakhsh \textit{et al}. (2021)} & {\small CO}$_{2}$ & {\small Foreign
			investment inflows} & {\small 40 Asian countries} & {\small 1996 - 2016} & 
		{\small FDI positive impact on CO2} & {\small Panel data} \\ 
		{\small Bidi and Jamil (2021)} & {\small CO}$_{2}$ & {\small GDP, Trade,
			FDI, financial and institutional Quality } & {\small Latin American,
			Caribbean, sub-Saharian} & {\small 2000 - 2018} & {\small Support} {\small %
			EKC except for sub-Saharian} & {\small Panel data econometric models} \\ 
		{\small Cheikh \textit{et al}. (2021)} & {\small CO}$_{2}$ & {\small Energy
			consumption, GDP Growth} & {\small MENA regions} & {\small 1980 - 2015} & 
		{\small Support EKC} & {\small Panel smooth transition modeling} \\ 
		{\small Khan and Eggoh (2021)} & {\small CO}$_{2}$ & {\small EG, trade,
			financial development, FDI.} & {\small 146 countries} & {\small 1990 - 2016}
		& {\small Support EKC} & {\small Smooth threshold Regression model} \\ 
		{\small Pandey and Mishra (2021)} & {\small SO}$_{2}${\small \ and NO}$_{2}$
		& {\small Net state domestic product, social expenditure } & {\small 21
			Indian states} & {\small 2001 - 2018} & {\small N-shaped EKC} & {\small %
			Panel unit root, cointegration, DOLS} \\ 
		{\small Wang \textit{et al}. (2021)} & {\small CO}$_{2}$ & {\small Urbanization} & 
		{\small OECD countries} & {\small 1960 - 2014} & {\small Support EKC} & 
		{\small Dynamic Panel ARDL model} \\ 
		{\small Beyene (in press) } & {\small Environment quality} & {\small Decomposed
			growth} & {\small 108 countries} & {\small 2000 - 2018} & {\small Not
			support EKC} & {\small Panel mean and quantile regressions} \\ 
		{\small Lee \textit{et al}. (2022)} & {\small 6 ecological footprints} & {\small %
			Tourism development} & {\small 99 countries} & {\small 2000 - 2017} & 
		{\small Support EKC} & {\small Quantile regression approach} \\
		{\small Zhong (2022)} & {\small SO}$_{2}${\small \ and CO}$_{2}$ & {\small %
			Income growth and inequality, industrial structure } & {\small China} & 
		{\small 2011 - 2015} & {\small Support EKC} & {\small Panel data methods} \\ 
		\hline
	\end{tabular}%
	\label{tab:panel}
\end{minipage}}\newline
\end{sidewaystable}

\subsubsection{Time series models}

Time series models are also very popular in the context of estimation of the EKC for a given country or territory. A large part of works using time series is based on single-equation models while some of them use multi-equation models; see the selected works reported in Table \ref{tab:TS}.

As mentioned above, given that the EKC is a long-run phenomena, model (\ref{eq:EKC}) could be estimated with the variables in levels after taking into account possible cointegration relationships between $e_{it}$ and $y_{it}$; see, for example, Waheed, Sarwar and Wei (2019), who survey several papers using Johansen cointegration tests and Umar \textit{et al}. (2020), who employ the combined cointegration and wavelet coherence approaches over the period 1971-2018 to explain the long run and causal effects of innovation, financial development and transportation infrastructure on CO$_2$ emissions in China. Alternatively, a very popular way of dealing with cointegration in the context of testing the EKC hypothesis is based on the ARDL bounds test of Pesaran, Shin and Smith (2001), which allows the possibility of including variables with different order of integration ($I(1)$ or $I(0)$), allowing for long- and short-run relationships and potential bi-directional relations; see the references reported in Table \ref{tab:TS} for several examples.

However, it is important to note that the possibility of non-stationarity of the variables involved in the estimation of the EKC opens the door to the estimation of spurious relations; see Wagner (2008), who shows that most studies estimating EKC gave spurious results. For example, Paruolo, Murphy and Janssens-Maenhout (2005) analyze the relation between income and emissions in all countries in the world over the period 1970-2008 based on a VEC model. They conclude that income and emissions seem to be driven by unrelated random walks plus drift.

Furthermore, it is important to point out that standard cointegration procedures may face problems when implemented in a non-linear context as that encountered when dealing with the EKC. M\"{u}ller-F\"{u}rstenberger and Wagner (2007) show that, if $\ln (y_{it})$, where $y_{it}$, is the \textit{per capita} GDP in region $i$ at time $t$, is a unit root process, its square is not an integrated process of order 1. Therefore, some previous findings obtained using standard unit-root and cointegration techniques both in panel or time series data may be questionable; see Perman and Stern (2003) for a discussion.

Recently, within the context of time-series regressions, some authors have shifted the focus away from the regression in levels in (\ref{eq:EKC}) to factor-augmented regressions in which the growth rate of GHG emissions in a given country is modelled as a function of factors extracted from a large (stationary) system of economic variables.\footnote{Note that short-run relations between emissions and the economy are also of interest because some policy instruments, such as emissions trading schemes, use economic incentives to control emissions over a short time spam, with the Regional Greenhouse Gas Initiative in US being an example.} Fosten (2019) fits the following diffusion index model to analyse the effects of economic activity on CO$_2$ emissions
\begin{equation}
\label{eq:difussion}
\varepsilon_t=\alpha+\delta^{\prime} \tilde{f}_t+\nu_t
\end{equation}
where $\varepsilon_t=\bigtriangleup log E_t$ with $E_t$ being yearly CO$_2$ emissions,\footnote{Fosten (2019) also considers emissions by source and non-linear specifications including polynomials of the factors.}. $\tilde{f}_t$ is the $r\times1$ vector of Principal Component (PC) factors extracted from a large set $X_t$ of macroeconomic variables and $\nu_t$ is assumed to be a white noise. Finally, $\alpha$ and $\delta=\left(\delta_1,...,\delta_r \right) ^{\prime}$ are parameters. Fosten (2019) also carries out a multi-resolution analysis of $y_t$ and $\tilde{f}_t$ based on discrete wavelet transformations, which decomposes the original monthly series into $J$ different time scales plus a smooth term. After extracting one single factor from $10$ selected variables of the data base described by McCracken and Ng (2016) and based on seasonally adjusted monthly data for U.S. observed from April 1978 to December 2018, Fosten (2019) concludes that emissions are not linked to economic activity in the short-run but there are strong linkages when looking at medium-run cycles of around one to three years.\footnote{Wavelet analysis has also been implemented by Jammazi and Aloui (2015), who analyse the relationship between energy consumption and economic growth using wavelet windowed cross-correlation for six oil exporting countries from the Gulf Cooperation Council (GCC) region, finding a bi-directional relationship. Kalmaz and Kirikkaleli (2019) also use a wavelet coherence approach to analyse the causal effects between CO$_{2}$ emissions and energy consumption, trade openness, urbanization and economic growth in Turkey. Finally, another study using wavelet techniques applied to US data is Raza, Shah and Sahrif (2019), who analyse several wavelet measures and conclude that in the short, medium and long-run energy consumption has positive influence over CO$_{2}$ emissions. They also findd unidirectional causality running from energy consumption to CO$_{2}$ emissions. 
}

Similarly, Mamipur, Yahoo and Jalalvandi (2019) extract factors from Iranian  systems of ``social'', ``economic'' and ``environmental'' variables and fit the following VAR model to the first differenced factors allowing for bi-directionality in the relations between the three systems
\begin{equation}
\label{eq:VAR}
\triangle \tilde{F}_t=\Phi \triangle \tilde{F}_{t-1} + a_t
\end{equation}
where $\tilde{F}_t=\left(\tilde{f}_{1t}, \tilde{f}_{2t}, \tilde{f}_{3t} \right) ^{\prime}$ is the vector of PC factors extracted from each of the systems.

In the same fashion, Bennedsen, Hillebrand and Koopman (2021) propose the following collapsed Structural Augmented Dynamic Factor Model (SADFM) for US CO$_2$ emissions 
\begin{equation}
\begin{pmatrix}
\varepsilon_t\\
x^{*}_t\\
\tilde{f}_t
\end{pmatrix}
=\begin{pmatrix}
\alpha\\
0\\
0
\end{pmatrix}
+ \begin{pmatrix}
\delta^{\prime}+\beta^{\prime}\Lambda^{*}\\
\Lambda^{*}\\
I_r
\end{pmatrix}
f_t + \begin{pmatrix}
\gamma^{\prime}\\
0\\
0
\end{pmatrix}
z_t + \begin{pmatrix}
u_t\\
v_t\\
e_t
\end{pmatrix}
\end{equation}
\begin{equation}
f_{t+1}=\Phi f_t + \eta_{t+1}
\end{equation}
where $\varepsilon_t=\bigtriangleup e_t$ with $e_t$ being the log-transformation of per capita CO$_2$ emissions observed yearly from 1960 to 2017, $x^{*}_t$ is a subvector of $X_t$, with $X_t$ containing the economic variables of special relevance, and $\tilde{f}_t$ is defined as in (\ref{eq:difussion}). Finally, $z_t$ are dummy variables that represent outliers; see Bennedsen, Hillebrand and Koopman (2021) for details about the estimation of the parameters, which is related with the two-step iterative procedure of Doz, Giannone and Reichlin (2012). They also consider a model for CO$_2$ emission rates depending on IP variables with time-varying parameters. In an out-of-sample exercise with forecast horizon $h=1$, their results favour the SADFM with $\delta=0$ over competing alternatives, and forecast decreases in emissions for 2018 and 2019.\footnote{Note that all variables in the SADFM are previously transformed to stationarity and, consequently, the diffusion index and SADFM models capture short-run relationships.}

Wagner, Grabarczyk and Hong (2020) deal with estimation and testing of the EKC for CO$_2$ emissions. The latter postulates an inverted U-shaped relationship between the level of economic development and CO$_2$ emissions, which in a time series setup implies a polynomial cointegration relationship between the emissions and the log of per capita gross domestic product. Typically, time series panels with small cross-sectional dimension for a limited number of countries are available, and it is quite natural to consider the country-specific EKCs as a system of seemingly unrelated relationships, including deterministic variables, integrated processes, and integer powers of integrated processes as explanatory variables. The authors develop two fully modified ordinary least squares type estimators for systems of seemingly unrelated cointegrating polynomial regressions, which account for serial correlation and for the endogeneity resulting from the existence of a cointegrating relationship. The paper derives the limiting distribution of the two estimators, in terms of a zero mean Gaussian mixture, and asymptotically chi-square Wald tests of linear restrictions, allowing to test for various forms of pooling. Poolability leads to estimation efficiency. Monte Carlo simulations illustrate the finite sample performance of the estimators and the tests. The empirical application considers a panel of 19 industrialized countries for the period 1870–2013 and finds evidence for quadratic cointegrating EKCs for a subset of six countries composed of Austria, Belgium, Finland, the Netherlands, Switzerland and the United Kingdom. The specific form of the EKC differs across these countries, but some group-wise pooling is supported by the tests, leading to sizeable parsimony and greater estimation efficiency. On the contrary, global pooling is strongly rejected.

\subsubsection{Regression models}

Due to the reduced sample and the heterogeneity of the countries included in the analysis, there are not so many studies using cross-sectional data as those using panel data or time series. For example, Magnani (2001) considers 152 countries in 1970, 1980 and 1990 and analyses the EKC\ relationship considering three pollutants (CO$_{2}$, SO$_{2}$ and nitrous oxide emissions), fitting model (\ref{eq:EKC}) in each of the three years. She finds different results depending on the degree of development of the economies. Similar analysis is done in Hill and Magnani (2002), using a sample of 156 countries and the same three previous pollutants. More recently, Chow and Li (2014) estimate the EKC relationship using 132 countries, considering a sample from 1992 to 2004. They estimate the EKC for each year separately and find strong support of the EKC. Finally, Atwi \textit{et al}. (2018) consider 182 countries and a sample from 1992 to 2011. They estimate the EKC relationship considering both a panel approach and a cross-section approach estimating model (\ref{eq:EKC}) for each year. They find support of the EKC when using panel data approach but not when considering cross-sectional regressions.

\begin{sidewaystable}
	\caption{Summary of selected works testing the EKC using time series methods.}
	\resizebox{0.6\textwidth}{!}{\begin{minipage}{\textwidth}
			\begin{tabular}{ccccccccccccc}
\hline
{\small Authors} & {\small Dependent Variable} & {\small Independent Variables} & {\small Countries} & {\small Sample} & {\small Conclusion} & {\small Method} \\
\hline
{\small Adejumo (2020)} & {\small CO}$_{2}$ & {\small EG} & {\small Nigeria} & {\small 1970 - 2014} & {\small Bidirectional relation environment and EG} & {\small ARDL} \\
{\small Koondhar \textit{et al}. (2020)} & {\small CO}$_{2},$ {\small NO}$_{2}$ & {\small Agricultural GDP, fertilizers, energy cons., grains} & {\small China} & {\small 1998 - 2018} & {\small Support EKC} & {\small ARDL} \\
{\small Pata and Aydin (2020)} & {\small CO$_{2}$} & {\small Hydropower energy cons., EG,} & {\small 6 hydropower cons.}  & {\small 1965 - 2016} & {\small Not support EKC} & {\small Fourier boostrapp ARDL\ .} \\
{\small Rahman (2020)} & {\small Ene. cons., CO}$_{2}$ & {\small EG, Pop. density, trade} & {\small India} & {\small 1971 - 2011} & {\small Different effects of EG on} {\small energy and CO}$_{2}$ & {\small ARDL} \\
{\small Rahman and Wu (2020)} & {\small CO}$_{2}$ & {\small Rew. Energy Cons., EG, trade, urbanizastion} & {\small Australia and Canada} & {\small 1960 - 2015} & {\small Different results LR and SR for each country} &  {\small ARDL, VECM} \\
{\small Rao and Yan (2020)} & {\small Different pollutants} & {\small EG} &  {\small Wuhan (China)} & {\small 1996 - 2015} & {\small Not support EKC for all pollutants} & {\small LARS/LASSO model} \\
{\small Rasool \textit{et al}. (2020)} & {\small CO}$_{2}$ & {\small Ene. cons., EG, inancial development} & {\small India} & {\small 1971 - 2014} & {\small Validate conventional EKC} & {\small ARDL, VECM} \\ 
{\small Sethi \textit{et al}. (2020)} & {\small CO}$_{2}$ & {\small Globalization, EG, Financial development} & {\small India} & {\small 1980 - 2015} & {\small Financial development not contribute to CO}$_{2}$ & {\small ARDL, VECM} \\ 
{\small Zmani and Ben-Salha\ (2020)} & {\small CO}$_{2}$ & {\small Pop., Affluence and Technology, urbanization, Energy Cons. } & {\small Gulf cooperation Council } & {\small 1980 - 2017} & {\small Determinants of CO}$_{2}${\small / volatility of EKC} & {\small PMG-ARDL\ approach} \\
{\small Mehmood (2021)} & {\small CO}$_{2}$ & {\small Globalization} & {\small Singapore} & {\small 1970 - 2014} & {\small Support EKC in Singapore} & {\small ARDL} \\ 
{\small Mele and Magazzino (2021)} & {\small PM}$_{2.5},${\small CO}$_{2},$ {\small NO}$_{2}$ & {\small EG} & {\small India } & {\small 1980 - 2018} & {\small Unidirectinal causality EG and pollution} & {\small Yamamoto test}\\
{\small Murshed (2021)} & {\small CO}$_{2},${\small NO}$_{2},${\small CH}$_{4}$ & {\small LPG cons. FDI, trade} & {\small 6 Asian Economies} & {\small 1980 - 2016} & {\small Support EKC India, Banglaseh, Sri Lanka, Bhutan} & {\small ARDL} \\
{\small Naeem \textit{et al}. (2021)} & {\small GDP per capita} & {\small Energy and non-Energy factors on EC} & {\small Pakistan } & {\small 1985 - 2018} & {\small Substitution of factors to improve EG} & {\small Ridge Regression}\\
{\small Ougan \textit{et al}. (2021)} & {\small CO}$_{2}$ & {\small Income pc increases, Income pc decreases} & {\small US} & {\small 1990.01 - 2019.07} & {\small EKC supported for decomposed model} & {\small ARDL} \\
{\small Pati\~{n}o \textit{et al}. (2021)} & {\small CO}$_{2}$ & {\small Sectorial decomposition of GDP} & {\small Colombia} & {\small 1971 - 2017} & {\small Additive and Multiplicative decomposition} & {\small DSGE models} \\
{\small Rehman \textit{et al}. (2021)} & {\small Economic progress} & {\small Urbanization, energy use, fossil fuel energy and CO}$_{2}$ & {\small China} & {\small 1975 - 2017} & {\small Asymmetric influence} & {\small Non linear ARDL}\\ 
{\small Sahbahz \textit{et al}. (2021)} & {\small CO}$_{2}$ & {\small Income pc, energy use, trade, oil price} & {\small India} & {\small 1980 - 2019} & {\small Asymmetric long term impact on CO}$_{2}$ & {\small Non linear ARDL}\\ 
{\small Solarin \textit{et al}. (2021)} & {\small Ecological footprint} & {\small EG, urbanization, FDI, trade} & {\small Nigeria} & {\small 1977 - 2016} & {\small Only EG contributes to disminish environment} & {\small ARDL} \\ 
{\small Villanthenkodath \textit{et al}. (2021)} & {\small CO}$_{2}$ & {\small EG and its components} & {\small India} & {\small 1971 - 2014} & {\small Conventional EKC hypothesis does not hold } & {\small ARDL} \\
{\small Xiang and Xu (2021)} & {\small GDP annual growth} & {\small Electricity prod. from different fonts, CO}$_{2}${\small emissions} & {\small China} & {\small 1995 - 2015} & {\small Ren. ene prod.contributes to increase in GDP} & {\small ARDL} \\
{\small Can Gen\c{c} \textit{et al}. (2022)} & {\small CO}$_{2}$ & {\small Volatility of EG} & {\small Turkey} & {\small 1980 - 2015} & {\small Support EKC} & {\small ARDL} \\
{\small Hashmi \textit{et al}. (2022)} & {\small CO}$_{2}$ & {\small GDP, Geopolitical risk and energy} & {\small Global} & {\small 1970 - 2015} & {\small Support EKC} & {\small ARDL} \\ 
{\small Kilinc-Ata and Likhachev (in press)} & {\small CO}$_{2}$ & {\small EG, ene.cons., pop., trade and financial development } & {\small Russia} & {\small 1990 - 2020} & {\small EKC supported} & {\small ARDL} \\
{\small Liu \textit{et al}. (2022)} & {\small Ecological footprint} & {\small tourism, EG, ene. cons., trade, FDI} & {\small Pakistan } & {\small 1980 - 2017} & {\small EKC depends on the independent variables} & {\small ARDL, Bayer and Hanck} \\
{\small Massagony and Budiono (in press)} & {\small CO}$_{2}$ & {\small Income, fossil and rene. ene cons.,forest restoration policies} & {\small Indonesia} & {\small 1970 - 2017} & {\small EKC supported} & {\small ARDL} \\
{\small Seri and de Juan (in press)} & {\small CO}$_{2}$ & {\small GDP and other control variables} & {\small Latin American countries} & {\small 1970 - 2018} & {\small Support EKC\ depending on country} & {\small ARDL VECM} \\
		\bottomrule
\end{tabular}
\label{tab:TS}
\end{minipage}}
					\begin{flushleft}
	{\scriptsize \textit{EG= Economic Growth, FDI=Foreign Direct Investment; Renew
			Ene: renewable energy; ene. cons. = energy consumption; Pop = population;
			pc = per capita;}.} 
\end{flushleft}
\end{sidewaystable}

\subsection{Main conclusions}

In general, the results described above are mixed and inconclusive with many open econometric issues; see, for example, the survey by Ul Husnain, Haider and Khan (2021). The empirical studies described in this section produce results that differ significantly depending on the period, the set of countries, the variables included, or the method of analysis. In general, regardless of the estimator implemented, most studies are based on annual data that is observed over a relatively short spam of time. Recently, many works estimate the EKC using single-equation time series models and the ARDL methodology.  Consequently, one could expect weak results when dealing with non-stationary data. Upon the basis of the studies reported in Tables \ref{tab:panel} and \ref{tab:TS}, we can conclude that there is not a strong support to the EKC. Evidence of an EKC is weak when we look at the worldwide level.

Finally, it is important to remark that most models in the literature focus on in-sample predictions of per capita CO$_2$ emissions. However, Auffhammer and Steinhauser (2012) argue that the interest is on out-of-sample forecasting of aggregate emissions. They argue the need for a shift in the emission forecasting literature toward criteria based on out-of-sample ability to predict total emissions versus the standard approach of selecting per capita models based on in-sample fit.

\section{Economic activity and climate change}
\label{section:Economy}

In this section, we survey the contributions looking at the (potential) effects of climate change on the economy and, in particular, on GDP. Several recent reviews are available. Tol (2009, 2014) describes the vast amount of research devoted to this topic. Focusing on GDP, he summarizes the results of several studies that use different methodologies to estimate the welfare cost (as percentage of GDP) of an increase on average temperature, where the main conclusions are that, on average, the effect is negative, being more pronounced in countries in Africa, Asia and South America. However, this effect is found to be positive for some European countries. Dell, Jones and Olken (2014) revise the literature of the effects of different climate variables (temperatures, precipitation and wind-storms) over economic outcomes, pointing out how the temporal dimension (exploited using panel data models) helps in identifying their effects. Apart from economic growth, Dell, Jones and Olken (2014) also review the effect on different outcomes as energy, labour productivity, industrial output and agriculture, among others. Burke, Hsiang and Miguel (2015) review the literature on the impacts of climate change on agricultural productivity and economic growth. The survey by Hsiang (2016) points out the main econometric issues in order to detect the effects of climate in social and economic outcomes. Hsiang and Kopp (2018) explain the physical science of climate change and presents evidence towards the rise of temperatures being due to human emissions. Finally, Nordhaus (2019) addresses climate change as externalities and describes the policy tools that are used in order to predict the effect of climate change in the economy. In this section, we summarize the main conclusions from these reviews and update them with the most recent contributions to the literature.

\subsection{GHG and Global Warming}

First, we survey works analysing the relationship between GHG emissions and GW. It is important to note that GW consists of two components: the greenhouse effect and the solar radiation effect, with both working in opposite directions. While an increase in GHG contributes to GW, more pollution causes an increase in aerosols (small particles that reflect and absorb sunlight in the atmosphere), so that less sunlight reaches the Earth. This phenomena is known as global dimming. Note that, in spite of its name, Magnus, Melenberg and Muris (2011) point out that global dimming is primarily a local (or regional) effect. Because of the dimming the Earth becomes cooler: the solar radiation effect. In practice, only the sum of the GHG and the solar radiation effects is observed. Magnus, Melenberg and Muris (2011) gather data from a large number of weather stations around the world for the period 1959–2002 and use dynamic panel data methods to separate both effects. They decompose the estimated temperature change of $0.73^{\circ}$C (averaged over the weather stations) into a greenhouse effect of $1.87^{\circ}$C, a solar radiation effect of $-1.09^{\circ}$C, and a small remainder term. Similarly, Phillips, Leirvick and Storelvmo (2020) use dynamic panel data methods to analyse the sensitivity of Earth's climate to a given increase in atmospheric GHG concentrations. Their analysis is based on spatio-temporal data of annual global surface temperatures, solar radiation and GHG concentrations observed over the last half century to 2010. They conclude that atmospheric aerosol effects masked approximately one-third of the continental warming due to the increasing GHG concentrations over this period; see also Storelvmo \textit{et al}. (2016) for similar conclusions. Petris and Hendry (2013) also conclude that there is a relationship between CO$_2$ and temperature supporting the anthropogenic forcing of climate change; see also Petris and Allen (2013), Estrada and Perron (2014), Stock (2019), Castle and Hendry (2020), Kim \textit{et al}. (2020) and Estrada, Kim and Perron (2021a, 2021b), among many others. Very recently, Chen, Gao and Vahid (in press) use the common features perspective and stablish that global temperatures and GHG share a common trend without conditioning on a particular type of trend and considering the possibility of endogeneity.

\subsection{Temperatures and economic activity}

Many studies analyse the effect of climate change on a particular sector or characteristic of the economy, which in the medium or long-run would affect growth. For instance, many papers focus on the effects of climate change on agriculture (Schlenker and Roberts, 2009, Burke and Emerick, 2016, Iizumi and Ramankutty, 2016, Hsiang \textit{et al}., 2017 and Gallic and Vermandel, 2020), exports (Jones and Olken, 2010), labour productivity (Seppänen, Fisk and Lei, 2006), energy demand (Auffhammer and Mansur, 2012), health (Deschênes and Greenstone, 2011, and Deryugina \textit{et al}., 2019), conflict (Miguel, Satyanath and Sergenti, 2004, Hsiang, Burke and Miguel, 2013, Hsiang and Burke, 2014) and crime and aggression (Miguel, 2005), among other outcomes; see also Carleton and Hsiang (2016) for various social and economic impacts of climate.

Earlier studies rely on cross-sectional regressions of several regions or countries observed in a given moment of time. The main finding of these early studies is the negative correlation between temperature and income per capita while there is no so clear connection between precipitation and income; see Dell \textit{et al}. (2014). Dell, Jones, and Olken (2009) show at country level that in 2000 an increase in $1^{\circ}$C translates into 8.5\% lower income per capita. Using municipal data for 12 American countries, they find that this drop is reduced to 1-2\% for municipal or province data, pointing out that other differences among countries besides climate change might be also responsible for the larger differences found among countries. However, even controlling for country fixed effects, Nordhaus (2006) concludes that 20\% of the income differences between rich industrial regions and  Africa can be due to geographical variables, among them, temperature and precipitation.

The same conclusion still holds in panel data analysis, the negative correlation between temperature and income per capita, although sometimes this is only found statistically significant for poorer countries. For instance, in a very well known work, Dell, Jones and Olken (2012), using a panel of countries observed from 1950 to 2003, conclude that there is a strong negative relationship between economic growth and warmer-than-average temperatures in poor countries while in rich countries this relationship is not significant. They also conclude that the effect of an increase of $1^{\circ}$C in annual temperature reduces income per capita growth rate by 1.4\%. Hsiang (2010) finds a more negative effect for Caribbean countries, with national output falling by 2.5\%. Moreover, when looking at temperature shocks, the effect turns out to be statistically significant only in the hottest season. However, some authors also find a negative effect of warm temperature on economic growth in rich countries. Burke, Hsiang and Miguel (2015) unify the seemingly contradictory results by accounting for non-linearity. They show that overall economic productivity is non-linear in temperature for all countries considered, with productivity picking at an annual average temperature of $13^{\circ}$C and declining strongly at higher temperatures. In spite of these results, there is no agreement in the literature on how the relation between GDP and temperature should be modelled. Newell, Prest and Sexton (2021) estimate more than 800 models!, depending on whether the effect of the temperature is linear or quadratic, GDP should be considered in levels or growth rates and the nature of some additional terms (trend, country fixed effects, time fixed effects, interaction fixed effects) to be included in the model, with mixed results in the outcomes pointing out the great uncertainty regarding model selection. Moreover, in an out of sample forecasting exercise, the effect of projections of unmitigated warming on GDP by 2100 also ranges from negative to positive depending on the model.

As regards precipitation, the conclusion is even less clear. Dell, Jones and Olken (2012) conclude that fluctuations in mean precipitation do not affect income per capita growth rate. However, Barrios, Bertinelli and Strobl (2010) found that, using standardized variables (the so-called weather anomalies) to account for climate fluctuations, higher rainfall is associated with higher growth rates in poor countries but not in rich ones; see, also, the literature review in their paper that suggests the inconclusive results regarding the effects of precipitation on growth. The reasons might be the type of precipitation variable used, the additional regressors or control variables introduced in the analysis and the sample used.

Recently, Kahn \textit{et al}. (2021) study the long-term impact of climate change on economic activity across countries using a panel model in which productivity is affected by deviations of temperature and precipitations from their long-term moving average historical norms. They analyse a panel of 174 countries observed annually from 1964 to 2014 and estimate the parameters using the half-panel jacknife fixed effects estimator of Chudick, Pesaran and Yang (2018), which allows to deal with biases due to temperature only being weakly exogenous. Also they allow for dynamics and feedback effects in the interconnections of climatic and macroeconomic variables, distinguishing between level and growth effects. They conclude that per-capita real output growth is adversely affected by persistent changes in temperature above and below its historical norm, but they do not obtain any statistically significant effect for changes in precipitation. Their empirical findings pertain to poor or rich, and hot or cold countries alike. However, the effects vary significantly across countries. One of the reasons that cold countries are also affected by climate change is the faster pace with which temperatures are rising in these regions when compared with that in hot countries. Also, the marginal effects of weather shocks are larger in low-income countries because they have lower capacity to deal with the consequences of climate change.  

Another branch of the literature with estimations relating climate variables and GDP growth is when climate variables are used as instruments of growth in first stage estimations using instrumental variables. Within this type of estimations, we also find mixed evidence of the relation between precipitations and economic growth. For instance, Miguel, Satayanath and Sergenti (2004) and Miguel and Satyanath (2011) found that variations in precipitations are positively related to GDP growth rate in panel studies although the relation seems to be weakening with time, while Burke and Leigh (2010) found negative correlation between precipitation and per capita income growth. In all these estimations the $R^2$ is quite low.

In the context of time series regressions, evidence for common breaks in temperature and radiative forcing could be consistent with the attribution of climate change to economic activity; see the editorial by Hillebrand, Petris and Proietti (2020).

\section{Economic effects of catastrophic weather phenomena}
\label{section:Disasters}

\begin{figure}[]
\includegraphics[trim={2cm 21cm 5cm 4cm},width=\textwidth]{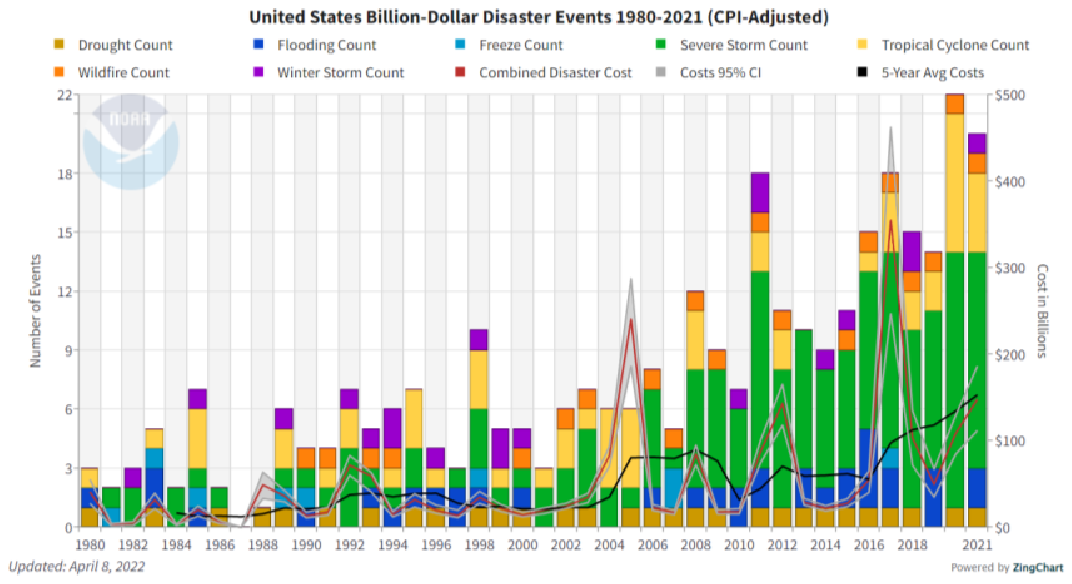}
\caption{US billion-dollar weather and climate disasters from 1980 to 2021 (CPI adjusted). The red line represents the combined disaster cost and the grey lines are 95\% confidence intervals. The black line represents 5-year average costs. Source: National Oceanic and Atmospheric Administration.}
\label{fig:billion}
\end{figure}

Climate change is likely to affect economies not only through warming, but also via an increase in extreme events like cold snaps and heat waves, floods and droughts, hurricanes, higher sea levels, and so on. Figure \ref{fig:billion} plots the number of annual weather and climate related disasters in US with a cost over one billion-dollar to illustrate this increase.\footnote{This plot was obtained from https://www.ncei.noaa.gov/access/monitoring/billions/time-series on 17th April 2022.} This figure also plots the direct costs adjusted for inflation using the Consumer Price Index (CPI), together with 95\% confidence intervals as well as the 5-year average costs. See Smith and Katz (2013), who describe the methodology used to estimate the US billion-dollar disaster loss, and conclude the costs are underestimated in roughly 10-15\%, and Smith and Matthews (2015) for a description of the Monte Carlo simulations used to obtain the confidence intervals. Figure \ref{fig:billion} suggests that increasing trends in both the annual frequency of billion-dollar events and in the annual aggregate loss from these events.\

Hoeppe (2016)  shows that direct losses are generally estimated using catastrophe models and measured using empirical data on losses. Most estimates of the economic impacts of natural disasters are based on regressions of aggregate variables (measured at the country level) on some measure of disasters, such as their number, the monetary damages or the number of fatalities.

The question of whether natural disasters affect economic growth is ultimately an empirical one; see the arguments in Cavallo \textit{et al}. (2013).  The economic losses of natural disasters have been increasing over the last few decades, with the number of natural disasters causing substantial losses increasing by a factor of three since 1980; see Pindyck (2013), who stresses the importance  of measuring the effects of climate change including the possibility of catastrophic outcomes, Schewe \textit{et al}. (2019), who warn against the under-estimation of the impacts of climate extremes by global models, and the reviews by Cavallo and Noy (2011), Botzen, Deschenes and Sanders (2019) and Hoeppe (2016) or the meta-analysis by Klomp and Valckx (2014).\footnote{Mittnik, Semmler and Haider (2020) analyse the link between CO$_2$ emissions and the frequency of climate related disasters.}

Most of this literature focus on short-run analysis. An exception is Cavallo \textit{et al}. (2013), who investigate the sign and size of the short- and long-run effects of large natural disasters on growth using event study with synthetic control groups. They conclude that, once one control for political changes, even extremely large disasters do not play any significant effect on economic growth. 

The literature considers different types of extreme events as inputs for the models in order to estimate their economic effect. As in Felbermayr \textit{et al}. (2022), we focus only on weather related events (windstorms, intense precipitation, droughts and cold spells) as input variables. 

When using {\em windstorms} as input variable and based on a sample covering the period 1950-2008, Nordhaus (2010) finds that, for the US, windstorms cost on average 0.07\% of GDP. However, Hsiang (2010) finds no average significant impact of cyclones on income using data of 28 Caribbean countries, although the effect can be positive in some sectors (construction) while negative in some others (tourism). Moreover, Strobl (2011), after proposing a novel hurricane destruction index, estimates the impact of hurricanes in US and concludes that they have at least a negative effect of 0.45 percentage points in coastal counties. The effect of windstorms in rich countries seems to be temporary and compensated through government transfers. For instance, at the county level in the US, Deryugina (2017) shows that 10 years after the hurricane, there is no effect on county income due direct and indirect government transfers in the form of unemployment insurance and public medical payments pointing out that the fiscal costs of natural disasters have been significantly underestimated due to these indirect transfers. In the particular case of hurricane Katrina in the US, Deryugina, Kawano and Levitt (2018) use a panel of tax return data and find that the effect of the hurricane on income was small and transitory, although it had a strong effect on where people live. Camargo and Hsiang (2016) survey the literature about the link between tropical cyclones and climate and also analyse their socio-economic impact, concluding that the direct and secondary economic impact of tropical cyclones is larger than previously thought. 

Another set of studies use night-light emissions as proxy for economic activity  and then the relation between night-light emision and economic growth, (see Bertinelli and Strobl, 2013 for the Caribbean, Elliot, Strobl and Sun, 2015, for China or Fellbermayr et al., 2022, for the whole globe) and then translate these results into GDP (for instance, Fellbermayr et al., 2022, use an elasticity of 0.37 of lights to GDP). The reasons for using night-light emissions are that they are available for the whole globe at the appropriate granularity (spatial resolution and time frequency), they also account for the informal economy and their measurement error is not correlated with the level of income per capita; see Felbrmayr \textit{et al}. (2022). The three analyses previously mentioned for the Caribbean, China and the whole globe, find negative effects of windstorms on night-light emission although Felbermayr \textit{et al}. (2022) find positive small spatial spillover effects.\footnote{Although the three studies look for spatial spillover effects, only Felbermayr \textit{et al}. (2022) include them directly in the model and find strong evidence of them with one lag.} 

Willner, Otto and Levermann (2018) show that the absence of enough adaptation to {\em pluvial floods} may provoke economic looses of around 17\% in the global trade network with China being the country with the most serious looses. However, Duan \textit{et al}. (2022) find that in China an additional day of heavy rainfall (more than 50 mm accumulated in 24 h) is associated with a 0.0092 percentage point increase in output. The rationale behind this contradicting result is as follows: Although there are clearly risks to floods and landslides from frequent precipitation in a short period, heavy rain helps economic development and the storage of water resources. There are also studies that use night-light emissions in order to see the economic effects of floods. Felbermayr \textit{et al}. (2022) use night-light emissions and, contrary to Duan \textit{et al}. (2021), find a negative relation between extreme precipitation and local income growth using regression fixed effects models. They also find that local income growth rebounds. Kokornik-Mina \textit{et al}. (2020) also relate floods with night-light emissions although they focus on cities. They find that economic activity is disrupted only briefly after a flood, returning to previous levels quickly.

Regarding {\em extreme temperatures}, there are analyses that focus on droughts or the effect of too high temperatures while other study cold spells. For the first one, Felbermayr \textit{et al}. (2022) built a Standardized Precipitation-Evapotranspiration Index, SPEI, and find that a unit increase of this index results in an increase of 0.8 percentage points in local night-light emission, with negative spillover effects resulting in a 0.44 decrease. 
For China, Duan \textit{et al}. (2022) study the effect of extreme heat (defined as the number of days that the mean temperature is higher than $32^{\circ}$C) and conclude that an additional day of extreme heat reduces GDP by 0.0006 percentage points with a heterogeneous impact across sectors, being the highest impact on industry, where an additional day of extreme heat implies that GDP loss increases 0.45\%; see Zhang \textit{et al.} (2018). 

For {\em cold spells}, Felbermayr \textit{et al}. (2022) find that their cold anomaly measure is negatively related to lights growth contemporaneously and in the next period. However, Duan \textit{et al}. (2022) study the effect of extreme cold for China (defined as the number of days that the mean temperature is lower than $-12^{\circ}$C) and do not find a significant relation. 

All in all, the main conclusion about the impact of natural disasters on income or GDP seems to be negative for all extreme input variables but heavy rain. However, this effect depends on the country being poor or rich. For the latter, the effects seem to be small and transitory, probably, due to the support of governments in form of  transfers or other indirect insurance.

\section{Conclusions}
\label{section:conclusions}

Relevant policy recommendations need to be based, among other considerations, on correct forecasts of the economic externalities caused by climate change. Consequently, robust measures of the interactions between environmental variables and the economy are crucial to formulate suitable economic development paths. In this paper, we survey over 250 econometric contributions to the literature on measuring these interactions over the last decade.

Credible quantitative modelling of climate change and its relations with economic activity is still in its very early stages. The steps forward taken during the last decades faces important econometric issues as, for example, endogeneity, non-stationarity, non-linearity and the importance of extreme events. The contributions in this area are not only crucial because of their implications for society, they are also a challenge for econometricians. Previous quantitative analysis of the externalities of climate change on the economy are limited in their conclusions because they only consider partial aspects of the environmental and economic variables and their relationships.

The global picture of the channels through which climate change and the economy may be related are well understood. First, economic activity causes an increase in energy consumption, which causes pollution. Larger pollution is related with climate change. Finally, climate change is itself affecting the economy.

However, the first issue faced to begin with is defining climate change. Climate is usually defined as the long-run average of weather in a given location (or globally) and climate change as the long-run variation in the joint probability distributions describing the state of the atmosphere, oceans, and fresh water including ice. In many studies analysing the relationship between climate change and economic activity, the measures used for the former are observations of weather variables, often temperatures and precipitations, obtained from weather stations. A majority of studies of climate change measure it by looking at temperatures and mainly to average temperatures.

With respect to the first channel, the nexus between energy consumption and economic growth has been analysed since long based mainly on panel data. Evidence on causality between energy consumption and economic growth still remains ambiguous.

There is also a huge and still growing literature on the nexus between economic activity and pollution. Most works focus on estimation of EKC based on either panel data or time series models. However, the time span considered in empirical studies is often relatively short while the variables considered are non-stationary. Consequently, the results are mixed and inconclusive.

With respect to the works looking at the relationship between economic activity and climate change, many papers focus on the effects of the latter on particular sectors while other, looking at a more aggregate level, find a negative correlation between temperature and income per capita (mainly in poorer countries). However, once more, evidence is mixed.

Finally, we survey the literature on the economic effects of catastrophic weather phenomena, which mainly focus on short-term effects. As before, the results are mixed, however, the empirical evidence seems to point to a negative relation between weather extreme events (but heavy pricipitation) and economic growth in poorer countries.

On top of describing what is known about the nexus between climate change and economic activity, this survey shows that a lot of work is required by econometricians in this very socially relevant topic. This analysis involves non-stationarity, endogeneity and non-linearity together with relatively short spans of data, being an important challenge for econometricians. Furthermore, the current approaches understated the large uncertainty around the economic effects of climate change, in particular they do not capture the downside risk. Very few empirical studies deal with the uncertainty problem around climate change. Pindyck (2014) and Convery and Wagner (2015) already discuss that standard climate-economy models fall victim of two important fallacies in dealing with uncertainty: by necessity, they focus on what is known and can be quantified and they convey a false sense of precision. According to our survey the same problem can be pointed out after several years.

\end{document}